\documentclass[twocolumn]{autart}  

\usepackage{graphicx}          
\usepackage{xcolor}
\usepackage{amsmath}
\usepackage{colortbl}
\usepackage{multirow}
\usepackage{booktabs}
\usepackage{bm}
\usepackage{tikzscale}
\usepackage{pgfplots}
\usepackage{tikz}
\usepackage{amssymb}
\usepackage{accents}
\usepackage{enumitem}
\usepackage{stfloats}

\newtheorem{assumption}{Assumption}
\newtheorem{proposition}{Proposition}
\newtheorem{lemma}{Lemma}

\newtheorem{corollary}{Corollary}
\newtheorem{theorem}{Theorem}

\newtheorem{remark}{Remark}

\newtheorem{definition}{Definition}
\pgfplotsset{compat=1.18}

\definecolor{red}{rgb}{0.9, 0.17, 0.31}
\definecolor{am}{rgb}{0.0, 0.6, 0.5}
\definecolor{cblue}{rgb}{0.16, 0.32, 0.75}

\def\cov{{\rm cov}}
\def\vec{{\rm vec}}
\def\svec{{\rm svec}}
\def\tr{{\rm tr}}
\def\row{{\rm row}}

\def\E{\mathbb{E}}
\def\N{\mathbb{N}}
\def\R{\mathbb{R}}
\def\S{\mathbb{S}}

\def\endproof{\begin{flushright} \vspace{-0.5cm} $\blacksquare$ \end{flushright}}

\def\e{\epsilon}

\newcommand{\X}{\ensuremath{\textcolor{am}{\bm{X}}}}
\newcommand{\SM}{\ensuremath{\textcolor{am}{\bm{S}}}}
\newcommand{\T}{\ensuremath{\textcolor{am}{\bm{T}}}}
\newcommand{\Nz}{\ensuremath{\textcolor{am}{\bm{N_0}}}}
\newcommand{\NzT}{\ensuremath{\textcolor{am}{\bm{N_0^T}}}}

\begin{document}

\begin{frontmatter}

\title{Linear Stochastic Systems with i.i.d. uncertainties: Exact Covariance Characterization, Stability Analysis and State-feedback Design  \thanksref{footnoteinfo}} 

\thanks[footnoteinfo]{This paper extends the works in \cite{Moussa2025} and \cite{Moussa2025b} accepted, respectively, in Automatica and IFAC WC 2026. Corresponding author K.~Moussa.}

 \author[UPHF,INSA]{Kaouther Moussa}\ead{kaouther.moussa@uphf.fr},    
 \author[LAAS]{Dimitri Peaucelle}\ead{peaucelle@laas.fr},
 \author[KyotoU]{Yohei Hosoe}\ead{hosoe@kuee.kyoto-u.ac.jp},
 \author[UGA]{Mirko Fiacchini}\ead{mirko.fiacchini@gipsa-lab.fr}

 \address[UPHF]{UPHF, CNRS, UMR 8201 - LAMIH, F-59313 Valenciennes, France}                                              
 \address[INSA]{INSA Hauts-de-France, F-59313, Valenciennes, France}           

 \address[LAAS]{LAAS-CNRS, Univ. Toulouse, CNRS, Toulouse, France} 
 
 \address[KyotoU]{Kyoto University, Kyoto, Japan} 

 \address[UGA]{Univ. Grenoble Alpes, CNRS, Grenoble INP, GIPSA-lab, 38000 Grenoble, France}


\begin{abstract} 
This paper studies linear discrete-time systems affected by independent and identically distributed (i.i.d.) multiplicative uncertainties and additive noise. It establishes the main links between covariance recursions, the spectral properties of associated Kronecker-based matrices, and mean-square stability, and exploits these links to derive tractable conditions for controller synthesis. We first derive a deterministic covariance recursion within the tube-based Stochastic Model Predictive Control (SMPC) framework using a Kronecker product based matrix augmentation. For linear stochastic systems with multiplicative uncertainty and
without additive noise, we show that the full-space matrix
representation arising from the covariance recursion has the same
spectral radius as its symmetric-space counterpart. Combined with the
existing symmetric-space characterization, this establishes that Schur
stability of the full-space augmented matrix is equivalent to
mean-square stability. For state-feedback design, we propose new sufficient Linear Matrix
Inequality (LMI) conditions that are numerically more tractable owing
to their reduced size compared with the conventional necessary and
sufficient conditions. Numerical tests illustrate the usefulness of the covariance characterization for recursively estimating the covariance without relying on sampling-based methods. We also assess the computational burden of the proposed LMI conditions and their conservatism relative to the necessary and sufficient ones.
\end{abstract}

\end{frontmatter}

\section{Introduction}
The class of stochastic systems provides a framework for describing dynamical systems affected by uncertainties of a probabilistic nature. Such uncertainties are inherent to many systems, including, for instance, systems modeling biological phenomena; see \cite{DelVecchio2018} and the references therein.

In contrast to robust approaches, which generally focus on the worst-case uncertainty realization, stochastic approaches aim to exploit probabilistic knowledge whenever it is available. This allows one to account for the likelihood of uncertainty realizations, for instance through chance-constrained formulations \cite{Cannon2011}. Such approaches have attracted considerable interest over the last decade. Among them, Stochastic Model Predictive Control (SMPC), see \cite{Mesbah2016} and the references therein, is closely related to the work presented in this paper.

The derivation of deterministic covariance recursions is essential for formulating identification and control problems for stochastic systems. Such recursions can be obtained straightforwardly in the case of additive noise. However, in the presence of multiplicative uncertainties, the main technical difficulty lies in decomposing terms of the form
\[
\E \left[A(\xi_k) P A(\xi_k)^T \right],
\]
where \(\xi_k\) denotes a random vector, \(A(\xi_k)\) is a random matrix, and \(P\) is a symmetric positive semidefinite matrix, representing, for instance, a covariance matrix. The decomposition of such terms consists in extracting the matrix \(P\) from the expectation.

This technical issue can be addressed using the vectorization operation and its connection with the Kronecker product, as done in \cite{Xing2022} for identification and in \cite{Moussa2025} for covariance stabilization, in the case of i.i.d. multiplicative uncertainties. The resulting augmented matrices are closely related to linear operators theory, and an equivalence can be established between their spectral properties and mean-square stability. In \cite{Coppens2023}, for affine-type stochastic uncertainties affecting the state and input matrices, the authors use the symmetric Kronecker product to derive a representation matrix restricted to the symmetric subspace. It is then shown that the spectral radius of this matrix is smaller than one if and only if the original system is mean-square stable.

In this paper, we show how the above mentioned expectation terms can be represented as deterministic linear operators through standard matrix decomposition. Furthermore,  we show that such operators have the same spectral radius when represented on the full space of real matrices and on the subspace of symmetric matrices. We then establish the link between these spectral properties and the necessary and sufficient conditions for mean-square stability presented in \cite{HOSOE2019}, for linear discrete-time stochastic systems with multiplicative uncertainties.

Finally, using the covariance recursions that we derive in the tube-based SMPC setting, we formulate sufficient LMI conditions ensuring the stability of the underlying dynamics. Although these conditions are only sufficient, they have the advantage of being lower-dimensional than standard necessary and sufficient conditions. The numerical tests presented in this paper illustrate their tractability, in particular for higher-dimensional systems.

\subsection*{Paper structure}
This paper is organized as follows.
Section~\ref{Section:Prob_statement} formulates the system under
consideration and the problems addressed in this paper.
Section~\ref{Section:Cov_charac} derives an exact deterministic
characterization of the covariance dynamics using a Kronecker-based
matrix representation.
Section~\ref{Section:background} recalls existing mean-square
stability and state-feedback design results.
Section~\ref{Section:Link_MSS} introduces the full-space and
symmetric-space representations of the second-moment operator and
establishes the equality of their spectral radii, thereby connecting
the full-space covariance recursion with the existing symmetric-space
characterization of mean-square stability.
Section~\ref{Section:Cov_Control} derives a reduced-size sufficient
LMI condition for the design of covariance-stabilizing state-feedback
gains.
Section~\ref{Section:Num_tests} presents numerical results illustrating
the exact covariance characterization and comparing the proposed
condition with the conventional necessary and sufficient condition in
terms of computational burden and conservatism.
Finally, Section~\ref{Section:Conclusion} summarizes the main results
and discusses possible directions for future work.

\subsection*{Notation}
$\mathbb{R}$ and $\mathbb{N}$ stand, respectively, for the sets of real and natural numbers. $\mathbb{R}^n$ and $\mathbb{R}^{n \times m}$ denote, respectively, the set of $n$-dimensional real vectors and the set of real matrices of dimension $n \times m$. We denote by $\mathbb{S}^{n}$ the set of symmetric matrices,  $\mathbb{S}^{n}_+$ and $\mathbb{S}^{n}_{++}$ denote, respectively, the sets of $n \times n$  symmetric positive semidefinite and definite matrices. The expectation of a random variable $x$ is denoted by $\mathbb{E}[x]$. Given a random vector $v$, $\cov(v)=\mathbb{E}\left[(v-\mathbb{E}[v])(v-\mathbb{E}[v])^T\right]$ stands for the covariance of $v$. If $v$ has zero mean, i.e., $\mathbb{E}[v]=0$, then $\cov(v)=\mathbb{E}\left[vv^T\right]$. The multivariate normal distribution with mean $\mu$ and covariance matrix $\Sigma$ is denoted by $\mathcal{N}(\mu,\Sigma)$. The notation $p\sim\mathcal{P}$ means that the random variable (or vector) $p$ follows the probability distribution $\mathcal{P}$. The Kronecker product is denoted by $\otimes$.

Given a square matrix $A \in \mathbb{R}^{n\times n}$, $A^{-1}$, $\tr(A)$, and $\rho(A)$ stand, respectively, for the inverse of $A$ (if nonsingular), its trace, and its spectral radius. The multiset consisting of the eigenvalues of $A$, including their algebraic multiplicities, is denoted by $\textnormal{mspec}(A)$. The zero matrix of appropriate dimensions is denoted by $0$, and the identity matrix of dimension $n$ is denoted by $I_n$.

Given a matrix $B \in \mathbb{R}^{n \times m}$, $\vec(B)$ and $\vec^{-1}$(B) stand, respectively, for the vectorization and inverse vectorization of $B$ (in column-major order).  The notation $\row(B)$ stands for the row-major order stacking operation. The transpose of a matrix $B$ is denoted by $B^T$. The notation $\|\cdot\|$ stands for the Euclidean norm for vectors and the induced spectral norm for matrices. For any matrix $X\in\mathbb{R}^{n\times m}$, $\|X\|_F := \sqrt{\tr(X^TX)}$ denotes its Frobenius norm. In particular, $\|\vec(X)\|=\|X\|_F$.

Given a square matrix $Q$, the notation $\{Q\}^{\mathcal{S}}$ stands for the symmetric matrix $\{Q\}^{\mathcal{S}} = Q+Q^T$, and $(\star)QB$ stands for the symmetric matrix $B^TQB$. Given a symmetric matrix $M \in \mathbb{S}^n$, $M \succ 0$ and $M \succeq 0$ mean that $M$ is positive definite and positive semidefinite, respectively. Given two symmetric matrices $A$ and $B$ of the same dimension, $A \prec B$ means that $B-A$ is positive definite. A matrix inequality of the type $I(\X)\prec 0$ is said to be a Linear Matrix Inequality (LMI) if $I(\X)$ is affine in the decision variables $\X$. In this paper, the decision variables are highlighted in \textcolor{am}{\textbf{bold green}}.

\section{Problem statement}\label{Section:Prob_statement}
Consider the following stochastic discrete-time linear system:
\begin{equation}
    x_{k+1} = A(\xi_k)x_k + B(\xi_k)u_k + w_k,
\label{Eq:sys_dyn}
\end{equation}
where \(x_k\in\mathbb{R}^{n}\) and \(u_k\in\mathbb{R}^{m}\), with \(n,m\in\mathbb{N}\), denote, respectively, the state and the control input. The random vectors \(\xi_k\in\mathbb{R}^{p}\), with \(p\in\mathbb{N}\), and \(w_k\in\mathbb{R}^{n}\) denote, respectively, the multiplicative uncertainty and the additive noise. The sequences \(\left(\xi_k\right)_{k\in\mathbb{N}}\) and \(\left(w_k\right)_{k\in\mathbb{N}}\) are assumed to be i.i.d. in time and mutually independent. The additive noise satisfies
$\mathbb{E}[w_k]=0$ and
$\mathbb{E}[w_kw_k^T]=W$,
where \(W\succ0\). The initial state \(x_0\in\mathbb{R}^{n}\) is assumed to be deterministic.

The entries of \(A(\xi_k)\) and \(B(\xi_k)\) are assumed to be square integrable, or equivalently,
\[
\mathbb{E}\left[\|A(\xi_k)\|_F^2\right]<\infty,
\qquad
\mathbb{E}\left[\|B(\xi_k)\|_F^2\right]<\infty.
\]
This assumption ensures that the second-order moments of the entries of the random matrices \(A(\xi_k)\) and \(B(\xi_k)\) are well defined, which is a basic requirement for characterizing and studying the associated covariance dynamics. 

Since \(\left(\xi_k\right)_{k\in\mathbb{N}}\) is i.i.d., all the random vectors \(\xi_k\),
\(k\in\mathbb{N}\), have the same distribution. Hence, the
second-order expectations considered throughout the paper do
not depend on \(k\), and can be equivalently written in terms
of either \(\xi_k\) or \(\xi_0\). No specific distributional
form is otherwise imposed on \(\xi_k\).

Considering that the control input depends only on the current state $x_k$, and under the   assumptions formulated above, \( A(\xi_k)\), \( B(\xi_k)\), and \(w_k\) are independent of \(x_k\) and \(u_k\) at time \(k\). Therefore, \(\mathbb{E}[ A(\xi_k)x_k] = \mathbb{E}[ A(\xi_k)]\,\mathbb{E}[x_k], \quad
\mathbb{E}[B(\xi_k)u_k] = \mathbb{E}[ B(\xi_k)]\,\mathbb{E}[u_k],\)
and
\(\mathbb{E}[x_k w_k^T] = \mathbb{E}[x_k]\mathbb{E}[w_k^T].\)

We consider, without loss of generality, the following decomposition:
\begin{equation}\label{Eq:mean_unc_decomp}
    A(\xi_k)=\bar{A}+\tilde{A}(\xi_k), \:\: B(\xi_k)=\bar{B}+\tilde{B}(\xi_k),
\end{equation}
such that $\mathbb{E}\left[ \tilde{A}(\xi_k)\right]=0$,  $\mathbb{E}\left[ \tilde{B}(\xi_k)\right]=0$ with $\bar{A} = \mathbb{E}\left[ A(\xi_k)\right]$ and $\bar{B} = \mathbb{E}\left[ B(\xi_k)\right]$. 

Under these assumptions, this paper addresses three main problems. \textbf{First}, we address the problem of deriving an exact deterministic characterization of the state covariance dynamics associated with system~\eqref{Eq:sys_dyn} in the tube-based SMPC framework. More specifically, we seek a finite-dimensional recursion that does not rely on sampling-based covariance propagation. This problem extends the work presented in \cite{Moussa2025}.

\textbf{Second}, we tackle the problem of linking the spectral properties of the Kronecker-based matrices arising from the covariance recursion to mean-square stability for systems without additive noise. We further investigate the connection between these spectral properties and the standard necessary and sufficient mean-square stability conditions for linear discrete-time systems with multiplicative uncertainties.

\textbf{Third}, we address the problem of deriving computationally tractable sufficient LMI conditions for covariance-stabilizing controller synthesis based on the obtained covariance recursion, with particular emphasis on conditions of reduced dimension. This problem extends the work presented in \cite{Moussa2025b}.

\section{Exact covariance characterization}\label{Section:Cov_charac}
This section addresses the exact deterministic characterization of the covariance dynamics associated with system~\eqref{Eq:sys_dyn}. Tube-based decomposition is one of the main approaches used to handle uncertainties in robust and stochastic MPC; see, for instance, \cite{Langson,Arcari2023}. It consists of separating the state into deterministic and uncertain components and designing a pre-stabilizing feedback to control the uncertain component. In stochastic MPC, this approach leads to tractable formulations in the presence of chance constraints. Deterministic covariance recursions have also been used for covariance propagation and identification; see, for example, \cite{Xing2022}, where the multiplicative uncertainties affecting the state and input matrices were assumed to be mutually independent. Motivated by SMPC frameworks, we derive here an exact covariance characterization in a tube-based setting without imposing this mutual-independence assumption. The resulting representation will subsequently be used to establish connections with standard mean-square stability notions.

Consider the decomposition
\begin{equation*}
    x_k=z_k+e_k,
\end{equation*}
where \(z_k\) denotes the nominal state and \(e_k\) denotes the uncertain error state. Using the decomposition in \eqref{Eq:mean_unc_decomp}, the nominal dynamics are defined as
\begin{equation}
    z_{k+1}=\bar A z_k+\bar Bv_k,
    \qquad
    z_0=x_0.
    \label{eq:z}
\end{equation}
Consequently, \(e_0=0\). We consider the affine control law
\begin{equation}
    u_k=Ke_k+v_k,
    \label{eq:tube_control}
\end{equation}
where \(K\) is a pre-stabilizing error-feedback gain and \(v_k\) is the nominal control input determined for instance by the MPC optimization problem. Combining \eqref{Eq:sys_dyn} and \eqref{eq:tube_control} gives
\begin{align}
    e_{k+1}
    ={}&
    (\bar A+\bar BK)e_k
    +\tilde A(\xi_k)x_k
    +\tilde B(\xi_k)u_k
    +w_k.
    \label{eq:e}
\end{align}

Since \(e_0=0\), the zero-mean properties of \(\tilde A(\xi_k)\), \(\tilde B(\xi_k)\), and \(w_k\), together with their independence from \(x_k\) and \(u_k\) at time \(k\), imply by induction that
\(
    \mathbb{E}[e_k]=0,
    \;
    \forall k\in\mathbb{N}.
\)
It follows that
\(
    z_k=\mathbb{E}[x_k]\) 
    and 
    \( \cov(e_k)=\mathbb{E}[e_ke_k^T].
\)
Moreover, since \(x_k=z_k+e_k\) and \(z_k=\mathbb{E}[x_k]\), then \(\cov(x_k)=\cov(e_k)\). 

Therefore, the covariance dynamics of system~\eqref{Eq:sys_dyn} can be characterized through those of the error state \(e_k\).

The following assumption concerns the asymptotic behavior of the nominal dynamics.

\begin{assumption}
\label{Ass:exp_stab}
The pair \((\bar A,\bar B)\) is stabilizable, and the nominal control policy \( v_k \) asymptotically stabilizes system~\eqref{eq:z}, so that
\( z_k\rightarrow0
    \quad\text{and}\quad
    v_k\rightarrow0 \)
as \(k\rightarrow\infty\).
\end{assumption}

We recall here the following standard identities that will be used in this section and throughout the paper; see \cite[Propositions~7.1.6 and~7.1.9]{Bernstein2009}. Let \(A\), \(B\), \(C\) and \(D\) be matrices of compatible dimensions. Then,
\begin{equation}
    \vec(ABC)
    =
    \left(C^T\otimes A\right)\vec(B).
    \label{Eq:Kron_prop_eq}
\end{equation}
Moreover, 
\begin{equation}
    (A\otimes B)(C\otimes D)
    =
    AC\otimes BD.
    \label{Eq:Kron_product_property}
\end{equation} 

We introduce the moment matrices
\begin{align}
    C_p^A
    &:=
    \mathbb{E}\!\left[
        \tilde A(\xi_k)\otimes\tilde A(\xi_k)
    \right],
    &
    C_p^B
    &:=
    \mathbb{E}\!\left[
        \tilde B(\xi_k)\otimes\tilde B(\xi_k)
    \right],
    \label{eq:moment_matrices_A_B}\\
    C_p^{BA}
    &:=
    \mathbb{E}\!\left[
        \tilde B(\xi_k)\otimes\tilde A(\xi_k)
    \right],
    &
    C_p^{AB}
    &:=
    \mathbb{E}\!\left[
        \tilde A(\xi_k)\otimes\tilde B(\xi_k)
    \right].
    \label{eq:moment_matrices_AB}
\end{align}

The following proposition extends the covariance characterization derived in \cite{Moussa2025} to systems in which multiplicative uncertainties affect both the state-transition and input matrices, without assuming mutual independence between these uncertainties.
\begin{proposition}
\label{prop:covariance_dynamics}
Let \(
    \epsilon_k
    :=
    \vec\!\left(\cov(e_k)\right),
    \qquad
    \omega
    :=
    \vec(W), \)
and define
\(\zeta_k^z
    :=
    \vec(z_kz_k^T),
    \qquad
    \zeta_k^v
    :=
    \vec(v_kv_k^T), \)
together with \(
    \zeta_k^{zv}
    :=
    \vec(z_kv_k^T),
    \qquad
    \zeta_k^{vz}
    :=
    \vec(v_kz_k^T). \)
Then, the covariance of the error state satisfies the exact deterministic recursion
\begin{align}
    \epsilon_{k+1}=& M(K) \epsilon_k + C_p^A \zeta_k^{z}  + C_p^B \zeta_k^{v}  + C_p^{BA} \zeta_k^{zv} \nonumber \\
    &+ C_p^{AB} \zeta_k^{vz} +\omega, 
    \label{eq:err_cov_dynamics}
\end{align}
where
\begin{align}
    M(K)
    :={}&
    (\bar A+\bar BK)\otimes(\bar A+\bar BK)
    +C_p^A
    +C_p^B(K\otimes K)
    \nonumber\\
    &+
    C_p^{BA}(K\otimes I_n)
    +C_p^{AB}(I_n\otimes K).
    \label{eq:Matrix_M}
\end{align}
The covariance matrix is recovered from
\(
    \cov(e_k)=\vec^{-1}(\epsilon_k).
\)
\end{proposition}
\paragraph*{Proof}
The derivation follows the same steps as the covariance characterization in \cite{Moussa2025}. In particular, one expands
\(\mathbb{E}[e_{k+1}e_{k+1}^T]\)
using \eqref{eq:e}. The terms involving single zero-mean matrices or the additive noise vanish by the independence and zero-mean assumptions. In contrast, the cross terms involving both \(\tilde A(\xi_k)\) and \(\tilde B(\xi_k)\) are retained, since these two random matrices are not assumed to be mutually independent. Applying the vectorization identity \eqref{Eq:Kron_prop_eq} and the mixed-product property \eqref{Eq:Kron_product_property} then yields the additional terms involving \(C_p^B\), \(C_p^{BA}\), and \(C_p^{AB}\).
\endproof

Proposition~\ref{prop:covariance_dynamics} provides a  deterministic representation of the covariance dynamics. The moment matrices in \eqref{eq:moment_matrices_A_B} and \eqref{eq:moment_matrices_AB} depend only on the second-order statistics of the multiplicative uncertainties and can therefore be computed independently of the covariance recursion. Their deterministic characterization in \eqref{eq:err_cov_dynamics} avoids the use of sampling-based covariance propagation. The main difficulty for controller synthesis arises from the nonlinear dependence of \(M(K)\) on \(K\), in particular through the term \(K\otimes K\).

\begin{remark}
\label{rem:symmetric_covariance_characterization}
The covariance recursion derived above is represented on the full
space \(\mathbb{R}^{n\times n}\), although covariance matrices belong
to the symmetric subspace \(\mathbb{S}^{n}\). The relation between the
spectral properties of this full-space representation and the standard
mean-square stability results is
investigated in Section~\ref{Section:Link_MSS}.

\end{remark}

The following corollary characterizes the convergence of the covariance recursion in \eqref{eq:err_cov_dynamics}.

\begin{corollary}
\label{Cor:Stability}
Let Assumption~\ref{Ass:exp_stab} hold. If there exists a gain \(K\) such that \(
    \rho\!\left(M(K)\right)<1, \)
then the covariance of the error state converges to
\begin{equation*}
    \lim_{k\rightarrow\infty}\cov(e_k)
    =
    \vec^{-1}\!\left(
        \left(I_{n^2}-M(K)\right)^{-1}\omega
    \right).
\end{equation*}
\end{corollary}
\paragraph*{Proof}
Assumption~\ref{Ass:exp_stab} implies that
\( \zeta_k^z,\;
    \zeta_k^v,\;
    \zeta_k^{zv}\) and 
    \(\zeta_k^{vz} \) converge to zero as  \(k\rightarrow\infty\).
The result then follows from the steady-state solution of the affine recursion~\eqref{eq:err_cov_dynamics}.
\endproof

\begin{remark}
Corollary~\ref{Cor:Stability} is related to the notion of mean-square boundedness considered, for instance, in \cite{Granzotto2024}. This notion refers to the exponential decrease of the expected value of the state squared norm  to a constant along the stochastic solutions, when an additive noise affects the dynamics.
\end{remark}

\section{Existing mean-square stability (MSS) and state-feedback design results}\label{Section:background}
We consider here the special case of system~\eqref{Eq:sys_dyn} without additive noise, given by
\begin{equation}\label{eq:sys_dyn_no_w}
x_{k+1}=A(\xi_k)x_k+B(\xi_k)u_k,
\end{equation}
under the same assumptions on \(\xi_k\), \(A(\xi_k)\), and \(B(\xi_k)\). Stability notions for this class of systems have been introduced in the literature, for example in \cite{Kozin1969}. In particular, in \cite{HOSOE2019}, the authors proved the equivalence between asymptotic, exponential and quadratic mean-square stability. We recall here the definition of mean-square asymptotic stability, which is related to the developments of this paper.

\begin{definition}\label{Def}
Let \((\xi_k)_{k\in\mathbb{N}}\) be an i.i.d. sequence of random vectors, and let \(A(\xi_k)\) be a random matrix whose entries are square integrable. The system
\begin{equation}\label{eq:sys_dyn_no_u}
  x_{k+1}=A(\xi_k)x_k  
\end{equation}
is mean-square asymptotically stable if, for each \(\varepsilon>0\), there exists \(\delta(\varepsilon)>0\) such that
\[
\|x_0\|\leq\delta(\varepsilon) \;
\Rightarrow \;
\mathbb{E}[\|x_k\|^2]\leq\varepsilon,
\quad \forall k\in\mathbb{N},
\]
and
\[
\mathbb{E}[\|x_k\|^2]\to0
\quad\text{as}\quad k\to\infty,
\]
for every deterministic initial condition \(x_0\in\mathbb{R}^n\).
\end{definition}

For the uncontrolled system \eqref{eq:sys_dyn_no_u}, quadratic stability is equivalent to the existence of \(P\in\mathbb{S}^n_{++}\) such that
\[
P-\mathbb{E}\!\left[A(\xi_0)^T P A(\xi_0)\right]\succ0.
\]

This condition takes the form of an expectation-based inequality,
as the decision variable \(P\) appears inside the expectation.
To obtain a deterministic representation, following
\cite{HOSOE2019}, consider a possibly rank-reduced factorization
\begin{equation*}
    \bar V ^T \bar V 
    =
    \mathbb{E}\left[
        (\star)
        \operatorname{row}\!\left(A(\xi_0)\right)
    \right],
\end{equation*}
where
\(\bar{V}\in\mathbb{R}^{r\times n^2}\), with
\(r\leq n^2\). A suitable block partitioning and reordering of
\(\bar{V}\) then yields a matrix
\(V \in\mathbb{R}^{n r\times n}\) such that
\begin{equation*}
    \mathbb{E}\left[
        A(\xi_0)^T P A(\xi_0)
    \right]
    =
    V ^T
    \left(P\otimes I_{r}\right)
   V.
\end{equation*}
Consequently, the necessary and sufficient stability condition can
be equivalently written as
\begin{equation}
    \exists P\in\mathbb{S}_{++}^{n}
    \quad\text{such that}\quad
    P-
    V^T
    \left(P\otimes I_{r}\right)
    V
    \succ 0.
    \label{LMI:existing_analysis}
\end{equation}
Although condition~\eqref{LMI:existing_analysis} resembles the standard discrete-time Lyapunov inequality, it is not a Lyapunov inequality directly associated with \(V\). In particular,  \(V\) is generally rectangular, and therefore the condition cannot be interpreted as a spectral-radius condition on  \(V\).

We now consider state-feedback design for
system~\eqref{eq:sys_dyn_no_w}. Under the state-feedback law
\(u_k=Fx_k\), a distinct expectation-based decomposition involving
both \(A(\xi_0)\) and \(B(\xi_0)\) is required. More precisely, it
is based on the decomposition of
\begin{equation}
    \mathbb{E}\left[
        (\star)
        \left[
            \operatorname{row}\!\left(A(\xi_0)\right),
            \operatorname{row}\!\left(B(\xi_0)\right)
        \right]
    \right].
    \label{Eq:expectation_matrix}
\end{equation}
The resulting necessary and sufficient synthesis condition is
recalled in the following theorem.

\begin{theorem}\cite{HOSOE2019}
    Let the sequence \(\left(\xi_k\right)_{k\in\mathbb{N}}\) be i.i.d. in time, and assume that the entries of \(A(\xi_k)\) and \(B(\xi_k)\) are square integrable. There exists a gain \(F\) such that the closed-loop system defined by \(x_{k+1}=\left(A(\xi_k)+B(\xi_k)F\right)x_k\) is quadratically stable in the second moment if and only if there exist \(X\in\mathbb{S}^{n}_{++}\) and \(Y\in\mathbb{R}^{m\times n}\) satisfying:
    \begin{equation}\label{LMI:Yohei2019}
        \begin{pmatrix}
            X & * \\
            \bar{G}_A^{'}X+\bar{G}_B^{'}Y & X\otimes I_{\bar{n}}
        \end{pmatrix}
        \succ 0,
    \end{equation}
where \(\bar{G}_A^{'}\in\R^{n\bar{n}\times n}\) and \(\bar{G}_B^{'}\in\R^{n\bar{n}\times m}\) result from the expectation-based decomposition of the matrix in (\ref{Eq:expectation_matrix}) as presented in \cite{HOSOE2019b}, with \(\bar{n}\leq n(n+m)\). In particular, \(F=YX^{-1}\) is such a stabilizing gain.
\label{Th:Yohei2019}
\end{theorem}

Since the three MSS notions, exponential, quadratic and asymptotic, are equivalent under the stated assumptions, condition \eqref{LMI:Yohei2019} also characterizes mean-square asymptotic stability of the closed-loop system. Note that this condition has dimension \(n+n\bar{n}\), which is at most \(n+n^2(n+m)\). The reader is referred to~\cite{HOSOE2019} for the decomposition details and decay-rate extensions. 

The recalled results provide necessary and sufficient conditions for mean-square
stability and state-feedback design. In particular, mean-square
stability admits a spectral characterization through the
second-moment operator represented on the symmetric subspace as presented in \cite{Coppens2023}.
However, the covariance characterization derived in
Section~\ref{Section:Cov_charac} is formulated using a full-space
matrix representation. The relation between these two
representations is established in the following section, yielding an equivalent spectral condition for stability analysis. Moreover,
the necessary and sufficient state-feedback condition
\eqref{LMI:Yohei2019} may involve an LMI of dimension up to
\(n+n^2(n+m)\), motivating the reduced-size sufficient condition
derived subsequently.

\section{Equivalence of full- and symmetric-space spectral representations}\label{Section:Link_MSS}
This section establishes the connection between the Kronecker-based matrices arising from the covariance recursion and standard mean-square stability notions. Starting from the autonomous system~\eqref{eq:sys_dyn_no_u}, we first show that its second-moment dynamics can be represented by a completely positive operator. We then establish that the full-space and symmetric-space representation matrices of this operator have the same spectral radius. This result allows us to characterize mean-square asymptotic stability directly in terms of the full-space Kronecker representation, which is the representation used in the covariance characterization of the previous section. Finally, we extend the result to the controlled system and show that the corresponding full-space representation coincides with the matrix \(M(K)\) introduced in~\eqref{eq:Matrix_M}.

\subsection{Symmetric-space representation and completely positive operators}

This subsection recalls the main algebraic tools used in the sequel. We first introduce the full-space and symmetric-space representations of linear matrix operators. We then recall the symmetric Kronecker product and its connection with completely positive operators and mean-square stability.

\paragraph*{Representations of linear matrix operators.}

We denote by \(\mathrm{SL}_n\) the set of linear operators
\(
    \mathcal{T}:\mathbb{R}^{n\times n}
    \rightarrow
    \mathbb{R}^{n\times n}
\)
that leave the symmetric subspace invariant, namely,
\(
    \mathcal{T}(\mathbb{S}^n)
    \subseteq
    \mathbb{S}^n.
\)

For any \(\mathcal{T}\in\mathrm{SL}_n\), its full-space representation matrix
\(\mathbf{M}_{\mathcal{T}}\in\mathbb{R}^{n^2\times n^2}\)
is uniquely defined by
\begin{equation*}
    \vec\!\left(\mathcal{T}(P)\right)
    =
    \mathbf{M}_{\mathcal{T}}\vec(P),
    \qquad
    P\in\mathbb{R}^{n\times n}.
\end{equation*}
To introduce the corresponding representation on the symmetric subspace, we recall the symmetric vectorization mapping introduced in \cite{De2002}.

\begin{definition}
For any symmetric matrix \(P\in\mathbb{S}^n\), the vector
\(\svec(P)\in\mathbb{R}^{d}\), with
\(d=\frac{n(n+1)}{2},\)
is defined as
\[
\begin{aligned}
    \svec(P)
    =
    \big(
    &p_{11},
    \sqrt{2}p_{21},
    \ldots,
    \sqrt{2}p_{n1},
    p_{22},
    \sqrt{2}p_{32},
    \ldots,\\
    &\sqrt{2}p_{n2},
    \ldots,
    p_{nn}
    \big)^T.
\end{aligned}
\]
\end{definition}
Let \(Q\in\mathbb{R}^{d\times n^2}\) denote the matrix associated with the above ordering of the entries of \(\svec(P)\), which satisfies \(QQ^T=I_d\) and, for every \(P\in\mathbb{S}^n\),
\begin{equation*}
    \svec(P)=Q\vec(P),
    \qquad
    \vec(P)=Q^T\svec(P).
\end{equation*}
The symmetric-space representation matrix
\(\mathbf{M}_{\mathcal{T}}^s\in\mathbb{R}^{d\times d}\)
is uniquely defined by
\begin{equation*}
    \svec\!\left(\mathcal{T}(P)\right)
    =
    \mathbf{M}_{\mathcal{T}}^s\svec(P),
    \qquad
    P\in\mathbb{S}^n.
\end{equation*}
Hence, we have
\begin{equation}
    \mathbf{M}_{\mathcal{T}}^s
    =
    Q\mathbf{M}_{\mathcal{T}}Q^T, \quad
    Q^T\mathbf{M}_{\mathcal{T}}^s
    =
    \mathbf{M}_{\mathcal{T}}Q^T.
\label{Eq:full_symmetric_relation}
\end{equation}

\paragraph*{Symmetric Kronecker product.}

The symmetric Kronecker product provides an explicit representation of operators acting on symmetric matrices.

\begin{definition}
For any \(G,H\in\mathbb{R}^{n\times n}\), the symmetric Kronecker product is defined as
\begin{equation*}
    G\otimes_s H
    :=
    \frac{1}{2}
    Q\left(G\otimes H+H\otimes G\right)Q^T.
\end{equation*}
\end{definition}

For every \(P\in\mathbb{S}^n\), the symmetric Kronecker for any \(G,H\in\R^{n\times n}\) can be viewed, by definition, as a mapping on a vector \(\svec\!\left(P\right)\):
\begin{equation*}
    (G\otimes_s H)\svec(P)
    =
    \frac{1}{2}
    \svec\!\left(
        HPG^T+GPH^T
    \right).
\end{equation*}
In particular, when \(G=H\),
\begin{equation}
    \svec\!\left(HPH^T\right)
    =
    (H\otimes_s H)\svec(P).
    \label{Eq:symmetric_Kronecker_quadratic}
\end{equation}
Consider now the linear operator 
\begin{equation}
    \mathcal{T}_A(P)=APA^T.
    \label{Eq:Lin_operator}
\end{equation}
Using \eqref{Eq:Kron_prop_eq} and
\eqref{Eq:symmetric_Kronecker_quadratic}, its full-space and symmetric-space representation matrices are, respectively,
\begin{equation}
    \mathbf{M}_{\mathcal{T}_A}
    =
    A\otimes A,
    \qquad
    \mathbf{M}_{\mathcal{T}_A}^s
    =
    A\otimes_s A.
    \label{Eq:single_operator_representations}
\end{equation}

In \cite[Theorem~4]{Blondel2005}, it is in particular shown that
\begin{equation*}
    \rho(A\otimes_s A)
    =
    \rho(A\otimes A).
\end{equation*}
In this section, we show that the equality between the spectral radii in the full and symmetric spaces also holds for operators whose representation matrices involve sums of Kronecker products.

\paragraph*{Completely positive operators.}

A class of linear operators of particular interest for covariance and second-moment dynamics is that of completely positive operators.

\begin{definition}
An operator \(\mathcal{T}\in\mathrm{SL}_n\) is said to be completely positive if there exist matrices
\(A_i\in\mathbb{R}^{n\times n}\), with \(i=1,\ldots,r\), such that
\begin{equation*}
    \mathcal{T}(P)
    =
    \sum_{i=1}^{r}A_iPA_i^T,
    \qquad
    P\in\mathbb{R}^{n\times n}.
\end{equation*}
The set of such operators is denoted by \(\mathrm{CP}_n\).
\end{definition}

Every operator \(\mathcal{T}\in\mathrm{CP}_n\) leaves \(\mathbb{S}^n\) invariant and maps \(\mathbb{S}_+^n\) into itself. By linearity of the standard and symmetric vectorizations, its full-space and symmetric-space representation matrices are, respectively,
\begin{equation}
    \mathbf{M}_{\mathcal{T}}
    =
    \sum_{i=1}^{r}A_i\otimes A_i,
    \qquad
    \mathbf{M}_{\mathcal{T}}^s
    =
    \sum_{i=1}^{r}A_i\otimes_s A_i.
    \label{Eq:CP_operator_representations}
\end{equation}

Building on the results of \cite{Kubrusly1985}, the work in
\cite{Coppens2023} establishes the equivalence between the Schur stability of
\(\mathbf{M}_{\mathcal{T}}^s\) and mean-square stability for stochastic linear dynamics with affine multiplicative noise.

The equality between the ordinary spectral radii of the two representation matrices in
\eqref{Eq:CP_operator_representations} will be established in
this section. This equality will subsequently be used to connect the Kronecker-based matrices arising from the covariance recursion with the mean-square stability conditions recalled in the previous section.

\subsection{Relation between the full- and symmetric-space representations}

Consider the autonomous system~\eqref{eq:sys_dyn_no_u},
under the assumptions stated in Section~\ref{Section:Prob_statement},  its second-moment dynamics satisfy
\begin{equation*}
    \mathbb{E}\!\left[x_{k+1}x_{k+1}^T\right]
    =
    \mathbb{E}\!\left[
        A(\xi_k)
        \mathbb{E}\!\left[x_kx_k^T\right]
        A(\xi_k)^T
    \right].
\end{equation*}

We introduce the linear operator
\begin{equation}
    \mathcal T(P)
    :=
    \mathbb{E}\!\left[
        A(\xi_k)PA(\xi_k)^T
    \right],
    \qquad
    P\in\mathbb R^{n \times n }.
    \label{Eq:Operator}
\end{equation}

The following lemma shows that \(\mathcal T\) admits a completely positive representation.

\begin{lemma}
\label{Lem:CP_decomposition}
There exist matrices \(A_i\in\mathbb R^{n\times n}\), with \(i=1,\ldots,r\) and \(r\leq n^2\), such that for every \( P \in \R^{n \times n}\) 
\begin{equation*}
    \mathcal T(P)
    =
    \sum_{i=1}^{r}A_iPA_i^T.
\end{equation*}
Hence, \(\mathcal T\in\mathrm{CP}_n\).
\end{lemma}

\paragraph*{Proof}
Consider the positive semidefinite matrix
\begin{equation*}
    \Gamma_A
    :=
    \mathbb{E}\!\left[
        (\star)
        \row\!\left(A(\xi_0)\right)
    \right].
\end{equation*}
Let
\[
    \Gamma_A=\bar V^T\bar V
\]
be a factorization of \(\Gamma_A\), and partition
\[
    \bar V^T
    =
    \begin{bmatrix}
        \bar V_1 & \bar V_2 & \cdots & \bar V_r
    \end{bmatrix},
\]
where \(\bar V_i\in\mathbb R^{n^2}\) and \(r=\operatorname{rank}(\Gamma_A)\leq n^2\). Define
\[
    U_i:=\vec^{-1}(\bar V_i) \in \R^{n \times n},
    \qquad
    A_i:=U_i^T,
\]
and 
\[
\mathcal{A} := \begin{bmatrix}
 A_1 &  A_2 & \cdots &  A_r
\end{bmatrix}.
\]
A direct block expansion of the factorization of \(\Gamma_A\) then gives
\[
    \mathbb{E}\!\left[
        A(\xi_k)PA(\xi_k)^T
    \right]
    = \mathcal{A} \left( I_r \otimes P\right) \mathcal{A}^T = 
    \sum_{i=1}^{r}A_iPA_i^T,
\]
for every \(P\in\mathbb R^{n \times n} \), which proves the result.
\endproof

Using the above introduced notation introduced, the full-space and symmetric-space representation matrices of \(\mathcal T\) are
\begin{align}
    \mathbf M_{\mathcal T}
    &=
    \sum_{i=1}^{r}A_i\otimes A_i
    =
    \mathbb{E}\!\left[
        A(\xi_k)\otimes A(\xi_k)
    \right],
    \nonumber \\
    \mathbf M_{\mathcal T}^{s}
    &=
    \sum_{i=1}^{r}A_i\otimes_s A_i
    =
    \mathbb{E}\!\left[
        A(\xi_k)\otimes_s A(\xi_k)
    \right].   \nonumber
\end{align}
Accordingly,
\begin{equation*}
    \vec\!\left(\mathcal T(P)\right)
    =
    \mathbf M_{\mathcal T}\vec(P),
    \qquad
    P\in\mathbb R^{n\times n},
\end{equation*}
and
\begin{equation*}
    \svec\!\left(\mathcal T(P)\right)
    =
    \mathbf M_{\mathcal T}^{s}\svec(P),
    \qquad
    P\in\mathbb S^n.
    \label{Eq:symmetric_action_T}
\end{equation*}

The following lemma extends the single-operator spectral comparison \cite[Theorem~4]{Blondel2005} recalled in  Section~\ref{Section:background} to completely positive operators whose representation matrices involve sums of Kronecker products.

\begin{lemma}
\label{Lem:rho_full_symmetric}
Let \(\mathcal T\in\mathrm{CP}_n\) be defined by
\[
    \mathcal T(P)
    =
    \sum_{i=1}^{r}A_iPA_i^T.
\]
Then,
\begin{equation*}
    \rho\!\left(\mathbf M_{\mathcal T}\right)
    =
    \rho\!\left(\mathbf M_{\mathcal T}^{s}\right).
\end{equation*}
\end{lemma}

\paragraph*{Proof}
From \eqref{Eq:full_symmetric_relation}, every eigenvalue of
\(\mathbf{M}_{\mathcal{T}}^s\) is also an eigenvalue of
\(\mathbf{M}_{\mathcal{T}}\). Indeed, if
\(\mathbf{M}_{\mathcal{T}}^s v = \lambda v\), then
\[
    \mathbf{M}_{\mathcal{T}} Q^T v
    =
    Q^T \mathbf{M}_{\mathcal{T}}^s v
    =
    \lambda Q^T v.
\]
Moreover, \(Q^T v \neq 0\) for \(v \neq 0\), since \(QQ^T=I\).
Therefore,
\[
    \rho\!\left(\mathbf{M}_{\mathcal{T}}^{s}\right)
    \leq
    \rho\!\left(\mathbf{M}_{\mathcal{T}}\right).
\]

To prove the reverse inequality, define
\[
    S_k
    :=
    \left\|
        \left(\mathbf{M}_{\mathcal{T}}^{s}\right)^k
    \right\|.
\]
For every \(k\in\mathbb{N}\), the \(k\)-th iterate of \(\mathcal{T}\) satisfies
\[
    \mathcal{T}^k(P)
    =
    \sum_{\alpha\in\{1,\ldots,r\}^k}
    A_{\alpha}P A_{\alpha}^T,
\]
where, for
\(
    \alpha=(\alpha_1,\ldots,\alpha_k)\in\{1,\ldots,r\}^k,
\)
we define
\(
    A_{\alpha}
    :=
    A_{\alpha_k}\cdots A_{\alpha_1}.
\)

Consequently, for every \(k\in\mathbb{N}\),
\begin{align*}
    \vec\!\left(\mathcal{T}^k(P)\right)
    &=
    \left(\mathbf{M}_{\mathcal{T}}\right)^k
    \vec(P),
    && P\in\mathbb{R}^{n\times n},\\
    \svec\!\left(\mathcal{T}^k(S)\right)
    &=
    \left(\mathbf{M}_{\mathcal{T}}^{s}\right)^k
    \svec(S),
    && S\in\mathbb{S}^{n}.
\end{align*}

For all \(u,v,p,q\in\mathbb{R}^n\), the Cauchy--Schwarz inequality gives
\begin{align}
    \left|
        p^T\mathcal{T}^k\!\left(uv^T\right)q
    \right|
    &=
    \left|
        \sum_{\alpha\in\{1,\ldots,r\}^k}
        \left(p^TA_{\alpha}u\right)
        \left(q^TA_{\alpha}v\right)
    \right|
    \nonumber\\
    &\leq
    \left(
        p^T\mathcal{T}^k\!\left(uu^T\right)p
    \right)^{1/2}
    \left(
        q^T\mathcal{T}^k\!\left(vv^T\right)q
    \right)^{1/2}. \nonumber
\end{align}

Summing this inequality over the canonical basis of \(\mathbb{R}^n\), and using
\[
    \operatorname{tr}(X)
    \leq
    \sqrt{n}\,\|X\|_F,
    \qquad
    X\in\mathbb{S}_+^n,
\]
yields
\[
    \left\|
        \mathcal{T}^k\!\left(uv^T\right)
    \right\|_F
    \leq
    \sqrt{n}\,S_k\|u\|\,\|v\|.
\]

Now let
\[
    P
    =
    \sum_{j=1}^{s}
    \sigma_j u_jv_j^T,
    \qquad
    s\leq n,
\]
be a singular-value decomposition of \(P \in \R^{n \times n}\), where
\[
    \sigma_j\geq0,
    \qquad
    \|u_j\|=\|v_j\|=1.
\]
By linearity of \(\mathcal{T}^k\),
\begin{align*}
    \left\|
        \mathcal{T}^k(P)
    \right\|_F
    &\leq
    \sum_{j=1}^{s}
    \sigma_j
    \left\|
        \mathcal{T}^k\!\left(u_jv_j^T\right)
    \right\|_F
    \\
    &\leq
    \sqrt{n}\,S_k
    \sum_{j=1}^{s}\sigma_j
    \\
    &\leq
    \sqrt{n}\,S_k\sqrt{s}
    \left(
        \sum_{j=1}^{s}\sigma_j^2
    \right)^{1/2}
    \\
    &\leq
    nS_k\|P\|_F.
\end{align*}
Consequently,
\[
    \left\|
        \left(\mathbf{M}_{\mathcal{T}}\right)^k
    \right\|
    \leq
    n
    \left\|
        \left(\mathbf{M}_{\mathcal{T}}^{s}\right)^k
    \right\|.
\]

Taking \(k\)-th roots and applying Gelfand's formula gives
\[
    \rho\!\left(\mathbf{M}_{\mathcal{T}}\right)
    \leq
    \rho\!\left(\mathbf{M}_{\mathcal{T}}^{s}\right),
\]
Combining both inequalities yields
\(
    \rho\!\left(\mathbf{M}_{\mathcal{T}}\right)
    =
    \rho\!\left(\mathbf{M}_{\mathcal{T}}^{s}\right).
\)
\endproof

The following proposition gathers equivalent spectral and convergence properties of the second-moment operator.

\begin{proposition}
\label{Prop:rho_conv}
For the operator \(\mathcal T\) defined in~\eqref{Eq:Operator}, the following assertions are equivalent:
\begin{enumerate}[label=(\alph*)]
    \item \(\rho\!\left(\mathbf M_{\mathcal T}\right)<1.\)
    \item \( \rho\!\left(\mathbf M_{\mathcal T}^{s}\right)<1.\)
    \item \( \left\| \left(\mathbf M_{\mathcal T}^{s}\right)^k \svec(P) \right\| \to0
        \quad\text{as}\quad k\to\infty,
        \quad
        \forall P\in\mathbb S^n.\)
    \item \(
        \left\|
            \left(\mathbf M_{\mathcal T}^{s}\right)^k
            \svec(vv^T)
        \right\|\to0
        \:\text{as}\quad k\to\infty,
        \quad
        \forall v\in\mathbb R^n.\)
    \item \(
        \left\|
            \left(\mathbf M_{\mathcal T}\right)^k
            \vec(P)
        \right\|
        \to0
        \quad\text{as}\quad k\to\infty,
        \quad
        \forall P\in\mathbb R^{n\times n}.\)
\end{enumerate}
\end{proposition}

\paragraph*{Proof}
The equivalence between \textit{(a)} and \textit{(b)} follows from Lemma~\ref{Lem:rho_full_symmetric}. The equivalences between \textit{(a)} and \textit{(e)} and between \textit{(b)} and \textit{(c)}  follow from the standard characterization of Schur stability in finite-dimensional spaces.

 The implication \textit{(c)}\(\Rightarrow\)\textit{(d)} is immediate. Conversely, for any \(P\in\mathbb S^n\), there exist \(\lambda_i \in \R\) and \(v_i\in\mathbb R^n\) such that
\[
    P
    =
    \sum_{i=1}^{q}\lambda_i v_iv_i^T.
\]
By linearity of \(\svec\),
\[
    \left\|
        \left(\mathbf M_{\mathcal T}^{s}\right)^k
        \svec(P)
    \right\|
    \leq
    \sum_{i=1}^{q}
    | \lambda_i |
    \left\|
        \left(\mathbf M_{\mathcal T}^{s}\right)^k
        \svec(v_iv_i^T)
    \right\|.
\]
Hence, \textit{(d)} implies \textit{(c)}, which completes the proof.
\endproof

\subsection{Link with mean-square stability}

The following theorem characterizes mean-square asymptotic stability through the full-space Kronecker-based representation matrix.

\begin{theorem}
\label{Thm:MSS_rhoM}
Let the sequence \(\left(\xi_k\right)_{k\in\mathbb{N}}\) be i.i.d. in time, and assume that the  entries of $A(\xi_k)$ are square integrable.  System~\eqref{eq:sys_dyn_no_u}, with deterministic initial condition \(x_0\), is mean-square asymptotically stable if and only if
\begin{equation*}
    \rho\!\left(\mathbf M_{\mathcal T}\right)<1.
\end{equation*}
\end{theorem}

\paragraph*{Proof}
Assume first that $\rho(\mathbf{M}_{\mathcal{T}})<1$. By Proposition~\ref{Prop:rho_conv}, for any deterministic initial condition $x_0\in\R^n$,
\[
\left\|\svec\!\left(\E[x_kx_k^T]\right)\right\|
=
\left\|\left(\mathbf M_{\mathcal T}^{s}\right)^k\svec(x_0x_0^T)\right\|\to 0
\; \text{as }k\to\infty.
\]
Therefore,
\[
\E[\|x_k\|^2]
=
\tr\!\left(\E[x_kx_k^T]\right)\to 0
\qquad\text{as }k\to\infty,
\]
which proves asymptotic convergence of the second moment.

It remains to prove Lyapunov stability in the second moment. Since $\rho(\mathbf{M}_{\mathcal{T}})<1$, Lemma~\ref{Lem:rho_full_symmetric} implies that $\rho(\mathbf{M}_{\mathcal{T}}^s) <1$, then the sequence $\left(\left(\mathbf M_{\mathcal T}^{s}\right)^k\right)_{k\geq 0}$ is bounded, and there exists a constant $c>0$ such that
\[
\|\left(\mathbf M_{\mathcal T}^{s}\right)^k\|\leq c,\qquad \forall k\in\N.
\]
Hence, for all $k\in\N$,
\[
\left\|\svec\!\left(\E[x_kx_k^T]\right)\right\|
=
\left\|\left(\mathbf M_{\mathcal T}^{s}\right)^k\svec(x_0x_0^T)\right\|
\leq
c\,\|\svec(x_0x_0^T)\|.
\]
Using $\|\svec(x_0x_0^T)\|=  \|\vec(x_0x_0^T)\| =  \|x_0x_0^T\|_F= \|x_0\|^2$, we obtain
\[
\left\|\E[x_kx_k^T]\right\|_F
\le
c\,\|x_0\|^2.
\]
Since $\E[x_kx_k^T]\in\S_+^n$, one has
\[
\E[\|x_k\|^2]
=
\tr\!\left(\E[x_kx_k^T]\right)
\le
\sqrt{n}\,\left\|\E[x_kx_k^T]\right\|_F
\le
\sqrt{n}\,c\,\|x_0\|^2.
\]
Therefore, there exists $\alpha>0$ such that
\[
\E[\|x_k\|^2]\le \alpha\|x_0\|^2,\qquad \forall k\in\N.
\]
Thus, for every $\epsilon>0$, choosing
\(
\delta(\epsilon)=\sqrt{\epsilon/\alpha}
\) ensures that
\[
\|x_0\|\leq \delta(\epsilon)
\quad\Longrightarrow\quad
\E[\|x_k\|^2]\leq \epsilon,\qquad \forall k\in\N.
\]
Hence, system~\eqref{eq:sys_dyn_no_u} is mean-square asymptotically stable.

Conversely, assume that system~\eqref{eq:sys_dyn_no_u} is mean-square asymptotically stable. Then, for every deterministic initial condition $x_0\in\R^n$,
\[
\E[\|x_k\|^2]\to 0
\qquad\text{as }k\to\infty.
\]
Since $\E[x_kx_k^T]\in\S_+^n$, one has
\[
\left\|\E[x_kx_k^T]\right\|_F
\leq
\tr\!\left(\E[x_kx_k^T]\right)
=
\E[\|x_k\|^2]\to 0.
\]
Therefore,
\[
\left\| \left(\mathbf{M}_{\mathcal{T}}^s\right)^k\svec(x_0x_0^T) \right\|
=
\left\|\svec\!\left(\E[x_kx_k^T]\right)\right\|\to 0.
\]
Hence, condition \textit{(d)} in Proposition~\ref{Prop:rho_conv} holds, which implies that $\rho(\mathbf{M}_{\mathcal{T}})<1$. This completes the proof.
\endproof

We now apply the previous theorem to the controlled system~\eqref{eq:sys_dyn_no_w} under the state-feedback law \( u_k=Kx_k, \) and define \begin{equation*} A_K(\xi_k) := A(\xi_k)+B(\xi_k)K. 
\end{equation*} The corresponding second-moment operator is \begin{equation*} \mathcal T_K(P) := \mathbb{E}\!\left[ A_K(\xi_k)P A_K(\xi_k)^T \right]. 
\end{equation*} 

\begin{proposition} \label{Prop:controlled_MSS_MK} Let the sequence \(\left(\xi_k\right)_{k\in\mathbb{N}}\) be i.i.d. in time, and assume that the entries of $A(\xi_k)$ and $B(\xi_k)$ are square integrable. The closed-loop system \[ x_{k+1} = \left( A(\xi_k)+B(\xi_k)K \right)x_k \] is mean-square asymptotically stable if and only if \begin{equation*} \rho\!\left(M(K)\right)<1, 
\end{equation*} where \(M(K)\) is defined in~\eqref{eq:Matrix_M}. \end{proposition}

\paragraph*{Proof} Noticing that the  full-space representation matrix of \(\mathcal T_K\), namely,  \begin{align} \mathbf M_{\mathcal T_K} &= \mathbb{E}\!\left[ A_K(\xi_k)\otimes A_K(\xi_k) \right] \nonumber\\ &= \mathbb{E}\!\left[ \left( A(\xi_k)+B(\xi_k)K \right) \otimes \left( A(\xi_k)+B(\xi_k)K \right) \right], \nonumber
\end{align} 
coincides with $M(K)$ and applying Theorem~\ref{Thm:MSS_rhoM} completes the proof. 
\endproof

 Proposition~\ref{Prop:controlled_MSS_MK} highlights in particular that the matrix \(M(K)\) obtained from the covariance recursion is also the full-space representation matrix governing the closed-loop second-moment dynamics. Indeed, 
 \begin{equation*} \vec\!\left( \mathbb{E}[x_{k+1}x_{k+1}^T] \right) = M(K) \vec\!\left( \mathbb{E}[x_kx_k^T] \right). 
 \end{equation*} Furthermore, \begin{align} \mathbb{E}\!\left[\|x_k\|^2\right] &= \mathbb{E}\!\left[ \operatorname{tr}(x_kx_k^T) \right] = \operatorname{tr}\!\left( \mathbb{E}[x_kx_k^T] \right) \nonumber\\ &= \operatorname{tr}\!\left( \cov(x_k) \right) + \operatorname{tr}\!\left( \mathbb{E}[x_k] \mathbb{E}[x_k]^T \right) \nonumber\\ &= \operatorname{tr}\!\left( \cov(x_k) \right) + \left\| \mathbb{E}[x_k] \right\|^2.\nonumber 
 \end{align} Hence, convergence of the second moment to zero is equivalent to convergence of both the covariance and the mean to zero. In the tube-based SMPC formulation of the previous section, \[ \mathbb{E}[x_k]=z_k, \qquad \cov(x_k)=\cov(e_k). \] Under Assumption~\ref{Ass:exp_stab}, the nominal mean \(z_k\) is asymptotically stabilized. Therefore, the condition \( \rho\!\left(M(K)\right)<1 \) ensures stability of the  covariance dynamics. In the absence of additive noise, this gives mean-square asymptotic stability. In the presence of additive noise, it ensures convergence of the covariance to the stationary matrix characterized in Corollary~\ref{Cor:Stability}. This spectral characterization provides the basis for the controller-design conditions developed in the next section.

Finally, note that since \(\cov(e_k)\in\mathbb{S}^n_+\), the recursion in \eqref{eq:err_cov_dynamics} can equivalently be represented on the symmetric subspace using the symmetric vectorization and symmetric Kronecker product introduced in this section. Such a representation reduces the dimension of the covariance state from \(n^2\) to \(n(n+1)/2\). The reduced expression is not reported explicitly because the controller-synthesis derivation in the next section relies on the full-space Kronecker representation, and the symmetric-space formulation does not provide a further reduction in the dimensions of the  LMI conditions derived subsequently. 

\section{Covariance-stabilizing controller design}\label{Section:Cov_Control}
The necessary and sufficient state-feedback condition recalled in
Section~\ref{Section:background} may involve an LMI of dimension up
to \(n+n^2(n+m)\). We therefore derive in this section a sufficient
condition of reduced size using the covariance representation
introduced in Section~\ref{Section:Cov_charac}. The following theorem provides a sufficient condition ensuring $\rho\left( M(K) \right) < 1$.

\begin{theorem}
Assume there exists $\X \in \S^{n^2}_{++}$, $\SM \in \R^{n \times n}$, $\T \in \R^{m \times n}$,  $\Nz \in \R^{(4n^2+nm) \times (3n^2+nm)}$ and the following holds 
\begin{equation}
  \left(\begin{array}{cc  cc   cc  cc  cc}
    \X&& 0 && 0  && 0&& 0\\
 0 && - \X && 0  && 0&&0\\
  0 && 0 && 0  && 0&&0\\
   0 && 0 && 0  && 0&&0\\
0 && 0 && 0  && 0&&0
\end{array}\right) \prec 
\left\{ N(\SM,\T)\NzT
\right\}^{\mathcal{S}}
\label{LMI:general}
\end{equation}  
where $N(\SM,\T)$ is defined in (\ref{Eq:N_matrix}). 
\begin{figure*}[!t]
\centering
\begin{equation}
N(\SM,\T):=  \left(\begin{array}{cc|  cc |  cc |  cc }
-I_{n} \otimes \SM && 0 && 0 && 0 \\
C_p^A \left(I_{n} \otimes \SM \right)   && (\bar A\SM+\bar B\T) \otimes I_n && C_p^B (\T \otimes I_m)+C_p^{AB} \left(\SM \otimes I_m \right) && C_p^{BA}(\T \otimes I_n) \\
I_n \otimes (\bar A\SM+\bar B\T) && -\SM \otimes I_n && 0&& 0  \\
I_{n} \otimes \T && 0 && -\SM \otimes I_m && 0\\
I_{n} \otimes \SM && 0 && 0 && - \SM \otimes I_n
\end{array}\right) 
\label{Eq:N_matrix}
\end{equation}
\end{figure*}
Then, $K = \T \SM^{-1}$ is such that $M(K)$ 
is Schur stable.
\end{theorem}
\paragraph*{Proof:}
Consider the dual system $\alpha^T_{k+1}=\alpha^T_k M(K)$ which is Schur stable if and only if $\epsilon_{k+1}=M(K)\epsilon_k$ is Schur stable. Let $\pi_k^{1T}=\alpha_k^T\left((\bar A+\bar B K) \otimes I_n\right)$, $\pi_k^{2T} = \alpha_k^T \left(C_p^B (K \otimes I_m) + C_p^{AB} \right)$, $\pi_k^{3T} = \alpha_k^T C_p^{BA} (K \otimes I_n)$ and $\eta_k^T=\left(\alpha_{k+1}^T,\alpha_k^T,\pi_k^{1T},\pi_k^{2T},\pi_k^{3T}\right)$. The dual system dynamics also read as the affine in $K$ descriptor form in (\ref{Eq:Descriptor_dual_dyn}). 
\begin{figure*}[!t]
\centering
\begin{equation}
\eta_k^T
\left(
\begin{array}{cc|  cc |  cc |  cc }
-I_{n^2} && 0 && 0 && 0 \\
C_p^A   && (\bar A+\bar B K) \otimes I_n && C_p^B (K \otimes I_m)+C_p^{AB} && C_p^{BA}(K \otimes I_n) \\
I_n \otimes (\bar A+\bar B K) && -I_{n^2} && 0&& 0  \\
I_{n} \otimes K && 0 && -I_{mn} && 0\\
I_{n^2} && 0 && 0 && - I_{n^2} 
\end{array}\right) = 0
\label{Eq:Descriptor_dual_dyn}
\end{equation}
\end{figure*}
Post-multiplying the equality in (\ref{Eq:Descriptor_dual_dyn}) by 
\[
\begin{pmatrix}
    I_{n} \otimes \SM & 0 &0& 0 \\
     0&   \SM \otimes I_n & 0 &0 \\
   0& 0 &  \SM \otimes I_{m}  &0\\
  0&  0 & 0 &   \SM \otimes I_{n}  \nonumber
\end{pmatrix},
\]
together with $\T=K\SM$ gives that the following holds along the trajectories
\(
\eta_k^T N(\SM,\T)=0.
\)
Therefore (\ref{LMI:general}) implies that 
\[
\eta_k^T\left(\begin{array}{cc  cc   cc  cc  cc}
    \X&& 0 && 0  && 0&& 0\\
 0 && - \X && 0  && 0&&0\\
  0 && 0 && 0  && 0&&0\\
   0 && 0 && 0  && 0&&0\\
0 && 0 && 0  && 0&&0
\end{array}\right)\eta_k
=
\alpha_{k+1}^T \X \alpha_{k+1}-\alpha_k^T \X \alpha_k
<0
\]
holds for the system. The matrix $\X$ is a Lyapunov certificate that proves stability of \( \alpha_{k+1}^T=\alpha_k^T M(K)\) and hence Schur stability of the matrix \( M(K) \).
\endproof 

The above proof is valid for any $\Nz \in \R^{(4n^2+nm) \times (3n^2+nm)}$ provided that (\ref{LMI:general}) holds. Moreover, for a fixed value of $\Nz$ the conditions are LMIs in the decision variables $\X$, $\SM$ and $\T$. In \cite{Moussa2025b}, we present a detailed discussion on the choice of such a matrix. In this paper, and for simplicity, we consider $\Nz$ as follows:
\[\NzT=
\left(
\begin{array}{cc  cc   cc  cc  cc}
    - I_{n^2} && 0 && 0 && 0 &&  I_{n^2} \\
       0 && 0 && - I_{n^2} && 0 && 0  \\
   0 && 0 && 0 &&  -I_{mn} && 0 \\
    0 && 0 &&0  && 0 && -I_{n^2} \\
\end{array}
\right)
\]

Finally, note that condition~(\ref{LMI:general}) has dimension $4n^2+nm$. For high-dimensional systems, this dimension is generally lower than that of the necessary and sufficient condition in (\ref{LMI:Yohei2019}), whose dimension is at most $n^2(n+m)+n$.

\section{Numerical tests and comparison}\label{Section:Num_tests}
Firstly, we present a numerical example to validate the exact characterization by recursively evaluating the theoretical covariance. The latter is then compared with the empirical covariance that is computed using Monte-Carlo trials. Secondly, we compare the computational complexity of the sufficient condition derived in (\ref{LMI:general})  to the necessary and sufficient condition in (\ref{LMI:Yohei2019}). Finally, we evaluate the conservatism of the derived sufficient condition in a specific setting.

\subsection{Exact characterization validation}

Consider the following dynamical system: 

\begin{align}\label{Eg:sys1}
x_{k+1} = 
&\begin{pmatrix}
1.2+\xi^A_{1k} & 1+\xi^A_{2k} \\
\xi^A_{3k} & 0.5+\xi^A_{4k}
\end{pmatrix}x_k+ \nonumber \\
&\hspace{1.5cm}\begin{pmatrix}
1 + \xi^B_{1k} & \xi^B_{2k} \\
\xi^B_{3k} & 1 + \xi^B_{4k}
\end{pmatrix}u_k+w_k,    
\end{align} 
with $W=I_2$. Let
\[
\xi_k^A
=
\left(
\xi^A_{1k},
\xi^A_{2k},
\xi^A_{3k},
\xi^A_{4k}
\right)^T,
\qquad
\xi_k^B
=
\left(
\xi^B_{1k},
\xi^B_{2k},
\xi^B_{3k},
\xi^B_{4k}
\right)^T.
\]
The matrices $\bar{A}$, $\bar{B}$, $\tilde{A}(\xi_k^A)$ and $\tilde{B}(\xi_k^B)$ are defined as follows:
\begin{equation*}
    \bar{A}= 
        \begin{pmatrix}
1.2 & 1 \\
0 & 0.5
\end{pmatrix},
\quad
\bar{B}= 
        \begin{pmatrix}
1 & 0 \\
0 & 1
\end{pmatrix},
\quad
    \tilde{A}(\xi_k^A)= 
        \begin{pmatrix}
\xi^A_{1k} & \xi^A_{2k} \\
\xi^A_{3k} & \xi^A_{4k}
\end{pmatrix},
\end{equation*}
\begin{equation*}
 \tilde{B}(\xi_k^B)=
 \begin{pmatrix}
\xi^B_{1k} & \xi^B_{2k} \\
\xi^B_{3k} & \xi^B_{4k}
\end{pmatrix}.
\end{equation*}
We consider that $\xi_k^A$ and $\xi_k^B$ are mutually independent and satisfy
\[
\xi_k^A\: \mathtt{\sim} \: \mathcal{N}\left(0,\Sigma\right),
\qquad
\xi_k^B\: \mathtt{\sim} \: \mathcal{N}\left(0,\Sigma\right),
\]
where
\begin{equation*}
\Sigma=
            \begin{pmatrix}
    3.94&    3.70&   3.72  &  4.08 \\
    3.70&    7.85&    6.95 &   7.12\\
    3.72&    6.95&    6.46 &   6.34\\
    4.08&  7.12  &    6.34 &   6.80
\end{pmatrix} \cdot 0.01,
\end{equation*}
resulting in the following matrices $C_p^A$ and $C_p^B$ after transformation: 
\begin{equation}\label{Eq:Cp}
C_p^A = C_p^B =
\begin{pmatrix}
    3.94&    3.70&   3.70 &   7.85\\
    3.72&    4.08&    6.95 &   7.12\\
    3.72&    6.95&    4.08 &   7.12\\
    6.46&    6.34&    6.34 &   6.80
\end{pmatrix} \cdot 0.01.
\end{equation}
Since $\xi_k^A$ and $\xi_k^B$ are independent and have zero mean, the matrices $C_p^{AB}$ and $C_p^{BA}$ are null. By solving the LMI condition in (\ref{LMI:general}), we obtain the following stabilizing gain
\begin{equation*}
    K= \begin{pmatrix}
   -0.8122  & -0.6824\\
   -0.0143 &  -0.3647
\end{pmatrix}.
\end{equation*}
We compute the evolution of the vectorization of the error covariance using the difference equation in~(\ref{eq:err_cov_dynamics}), as well as the empirical covariance based on $N$ Monte Carlo tests. We denote by $\e_{ij}^{th}$ and $\e_{ij}^{em}$, respectively, the theoretical and the empirical elements of the error covariance matrix, for $i,j\in \{1,2\}$. $x_1$ and $x_2$ stand for the first and second elements of $x_k$ (uncertain state), while $z_1$ and $z_2$ stand for the first and the second elements of $z_k$ (nominal state).  
 
\begin{figure}
    \centering
\includegraphics[scale=0.55]{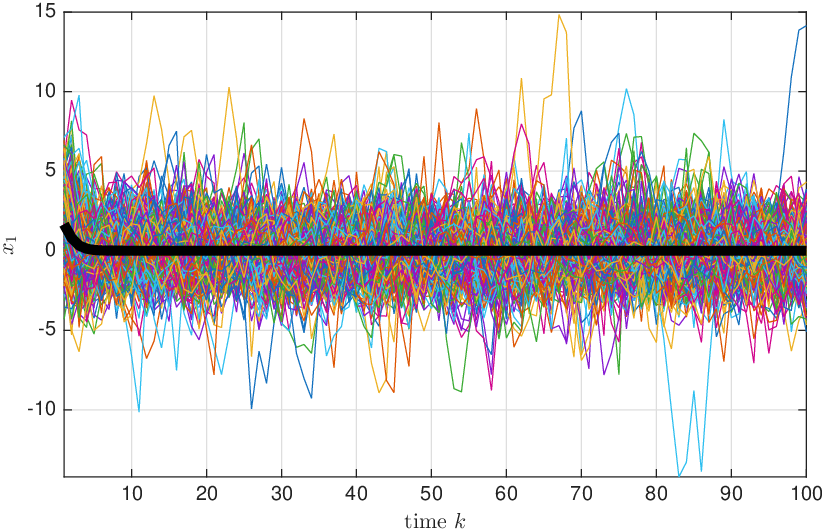}
    \caption{Trajectories of $x_1$ and its corresponding nominal state $z_1$ (in bold black).}
    \label{fig:x1_traj_N500}
\end{figure}

\begin{figure}
    \centering
    \includegraphics[scale=0.55]{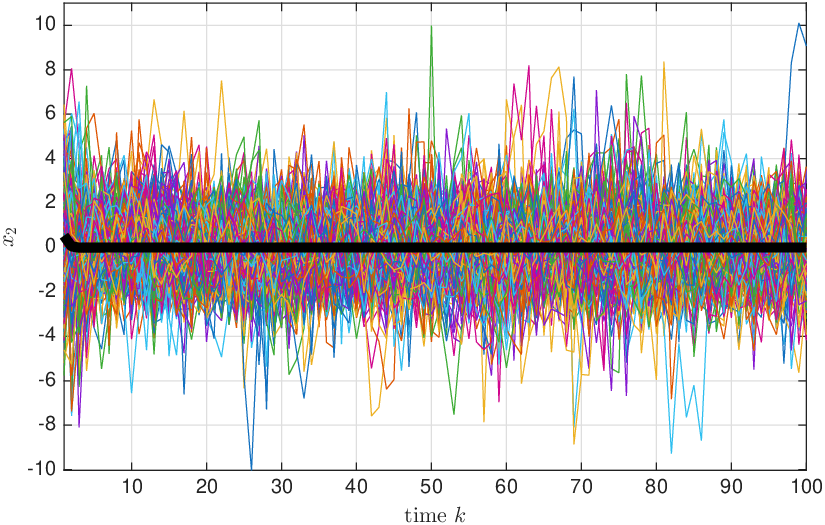}
    \caption{Trajectories of $x_2$ and its corresponding nominal state $z_2$ (in bold black).}
    \label{fig:x2_traj_N500}
\end{figure}

Fig.~\ref{fig:x1_traj_N500}~and~\ref{fig:x2_traj_N500} show the uncertain trajectories for 500 Monte Carlo tests, with their respective nominal trajectories that are driven to 0. Note that in this example, the same stabilizing gain $K$ has been used for the nominal dynamics, \textit{i.e.} $v_k = K z_k$. In the case of a stochastic MPC implementation, $v_k$ should be designed by a deterministic MPC strategy.
 
\begin{figure}
    \centering
    \includegraphics[width=0.8\linewidth]{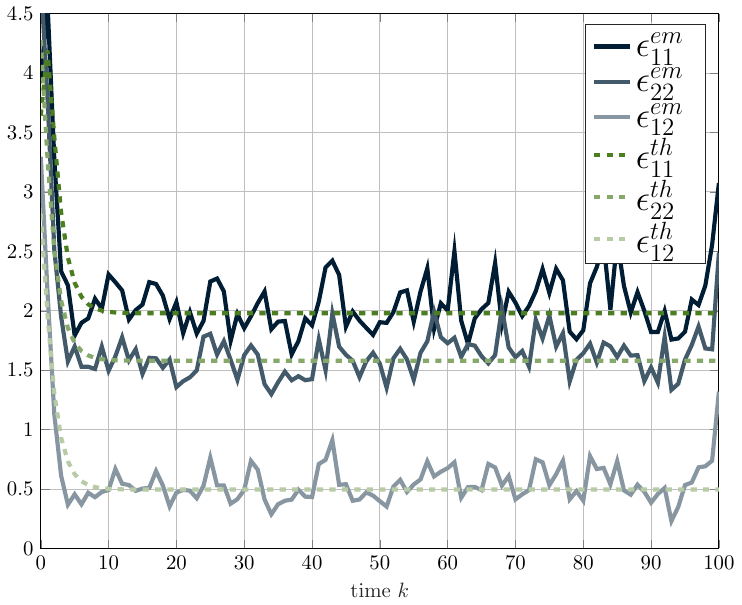}
    \caption{Theoretical and empirical error covariance evolution related to system~(\ref{Eg:sys1}) with $C_p^A$ and $C_p^B$ as in (\ref{Eq:Cp}), evaluated with $N=500$ Monte Carlo trials.}
    \label{fig:cov_traj_N500}
\end{figure}

\begin{figure}
    \centering
    \includegraphics[width=0.8\linewidth]{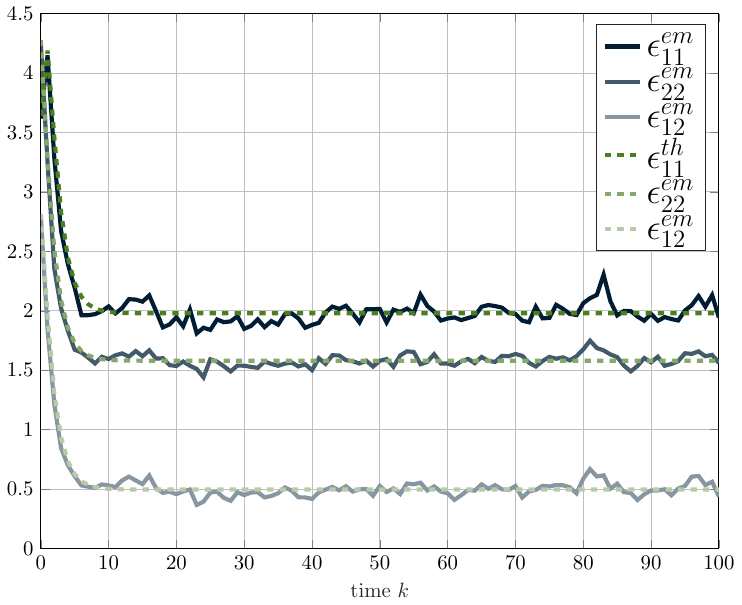}
    \caption{Theoretical and empirical error covariance evolution related to system~(\ref{Eg:sys1}) with $C_p^A$ and $C_p^B$ as in (\ref{Eq:Cp}), evaluated with $N=5000$ Monte Carlo trials.}
    \label{fig:cov_traj_N5000}
\end{figure}

Fig.~\ref{fig:cov_traj_N500}~and~\ref{fig:cov_traj_N5000} show both theoretical and empirical covariance entries, evaluated using Monte Carlo tests $N=500$ and $N=5000$, successively. We can see that the empirical error covariance matches the theoretical one. Furthermore, we can notice that depending on the chosen $N$, the empirical covariance might exhibit a noisy behavior. Therefore, $N$ has to be increased in order to have an accurate approximation of the covariance, which highlights the relevance of having a theoretical exact characterization for such systems. Note that both the empirical and theoretical covariance matrices converge to the following matrix, as stated in Corollary~\ref{Cor:Stability},
\begin{equation*}
    \vec^{-1}\left(\left( I-M(K)\right)^{-1} \vec \left(W\right)\right) =
    \begin{pmatrix}
1.98  &  0.50 \\
    0.50   & 1.58
\end{pmatrix}.
\label{Ex:cover_err}
\end{equation*}

\subsection{Computational-time comparison}
\begin{figure}
    \centering
    \includegraphics[width=0.8\linewidth]{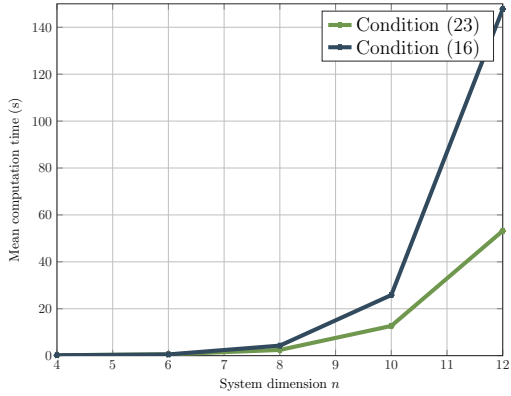}
    \caption{Numerical tests with stochastic input and state transition matrices}
    \label{fig:both_stoc}
\end{figure}

In this section, we compare the computational complexity of the sufficient condition derived in (\ref{LMI:general}) to the necessary and sufficient condition in (\ref{LMI:Yohei2019}) from \cite{HOSOE2019}. Since the asymptotic stability notion is only related to systems without additive noise, let us consider systems as defined in~(\ref{eq:sys_dyn_no_w}). The following setting has been considered: 
\begin{itemize}
    \item Random nominal matrices $\bar{A}$ of size $n$ were generated such that both conditions were feasible and $\bar{B}=I_n$. 
    \item The entries of $\tilde{A}(\xi_k)$ and $\tilde{B}(\xi_k)$ are assumed to follow normal distributions with 0 mean and variance 0.05.  
    \item The mean time was evaluated using 100 tests. 
\end{itemize}
Fig.~\ref{fig:both_stoc} shows that condition (\ref{LMI:general}), of size $4n^2+nm$, has a lower computational time compared to condition (\ref{LMI:Yohei2019}), the latter being of size $n^2(n+m)+n$ in the considered example, this difference increases as $n$ grows. 

\subsection{Conservatism assessment}
In order to evaluate the conservatism of the derived sufficient condition with respect to the necessary and sufficient one, let us consider systems as defined in~(\ref{eq:sys_dyn_no_w}), where the stochastic matrices $\tilde{A}(\xi_k)$ and $\tilde{B}(\xi_k)$ entries follow the normal distribution defined as $\mathcal{N}\left(0,\sigma^2\right)$. For each setting, the variance is increased until the conditions are no more feasible, then, this maximal variance is denoted $\sigma_{max}^2$. We consider the following settings for system~(\ref{eq:sys_dyn_no_w}), with $\bar{B}=I_2$: 
\begin{equation*}
   \bm{\Gamma_1} \: :\: \bar{A} = 
    \begin{pmatrix}
    3.5 &0.2\\
    1.1 & 1.5 
\end{pmatrix}, \: \text{mspec}(\bar{A}) = \{     3.6045, 1.3955\}, \:     
\end{equation*}
\begin{equation*}
   \bm{\Gamma_2} \: :\:  \bar{A} = 
    \begin{pmatrix}
    1.2 &0.2\\
    1.1 & 1.2 
\end{pmatrix}, \: \text{mspec}(\bar{A}) = \{     1.6690, 0.7310\}, \: 
\end{equation*}
\begin{equation*}
   \bm{\Gamma_3} \: :\:  \bar{A} = 
    \begin{pmatrix}
    1.4 &0.1\\
    0.2 & 1.2 
\end{pmatrix}, \: \text{mspec}(\bar{A}) = \{         1.4732,   1.1268\}.
\end{equation*}

\begin{table}
\centering
\renewcommand{\arraystretch}{1.3}
\arrayrulecolor{black}

\begin{tabular}{
    >{\columncolor{gray!25}}c |
    >{\columncolor{gray!25}}c >{\columncolor{gray!10}}c |
    >{\columncolor{gray!25}}c >{\columncolor{gray!10}}c
}
\rowcolor{gray!0}
 & \multicolumn{2}{c|}{\scriptsize \cellcolor{gray!50}Condition (\ref{LMI:general})} & \multicolumn{2}{c}{\scriptsize \cellcolor{gray!50}Condition (\ref{LMI:Yohei2019})} \\
\rowcolor{gray!0}
 &\cellcolor{gray!25} \tiny{ $\bm{\sigma^2_{max}}$} &\cellcolor{gray!25} \tiny{$\bm{K}$} & \cellcolor{gray!25}\tiny{$\bm{\sigma^2_{max}}$} &\cellcolor{gray!25} \tiny{$\bm{K}$} \\
\hline
\rowcolor{gray!10}
\scriptsize{\textbf{$\bm{\Gamma_1}$}} & -- & --
     & \tiny{0.06} &  \tiny  \scalebox{0.8}{$ \left(\hspace{2pt} \begin{matrix} 
               -3.2893  & -0.2364\\
   -0.9356   &-1.1317
        \end{matrix} \right)$} \\
\rowcolor{gray!10}
\scriptsize{\textbf{$\bm{\Gamma_2}$}} & \tiny{0.09} &  \tiny \scalebox{0.8}{$\left(\hspace{2pt} \begin{matrix}
-0.9824&-0.2003\\
-0.8038&-0.8422
        \end{matrix} \right)$} & \tiny{0.19} & \tiny \scalebox{0.8}{$       \left(\hspace{2pt} \begin{matrix}
   -0.9860  & -0.2575\\
   -0.8227  & -0.7944
        \end{matrix} \right)$} \\
\rowcolor{gray!10}
\scriptsize{\textbf{$\bm{\Gamma_3}$}} & \tiny{0.16} & \tiny\scalebox{0.8}{$     \left(\hspace{2pt}   \begin{matrix}
-1.0001&-0.0811\\
-0.1512&-0.8405
        \end{matrix} \right)$} & \tiny{0.22} & \tiny \scalebox{0.8}{$ \left(\hspace{2pt}\begin{matrix}
   -1.0133  & -0.0874\\
   -0.1585  & -0.8449
        \end{matrix} \right)$} \\
\end{tabular}
\caption{Comparison of conditions (\ref{LMI:general}) and (\ref{LMI:Yohei2019})}
\label{Table:both_stoc}
\end{table}

Table~\ref{Table:both_stoc} shows the maximal variance $\sigma_{max}^2$ for which the conditions in (\ref{LMI:general}), and (\ref{LMI:Yohei2019}) are feasible, for different systems; in addition to the computed pre-stabilizing gain $K$. The boxes with the sign -- correspond to cases where the problem was not feasible, even for very low variances. The presented results highlight the fact that the maximal variances for which the condition in (\ref{LMI:general}) remains feasible are lower than those related to condition (\ref{LMI:Yohei2019}). This difference in the maximal variances seems to increase with the instability of the matrix $\bar{A}$, measured by the deviation of its eigenvalues from the unit circle. This highlights the fact that the conservatism of the sufficient condition in (\ref{LMI:general}) depends on the system under study.

\section{Discussion \& Conclusion}\label{Section:Conclusion}
This paper presented an exact deterministic characterization of the covariance dynamics of discrete-time linear systems affected by i.i.d. additive and multiplicative uncertainties, extending our previous work in \cite{Moussa2025}. It also established the connection between the spectral properties of the resulting covariance and second-moment dynamics and mean-square asymptotic stability of stochastic linear discrete-time systems.

Moreover, a new sufficient condition for the design of stabilizing feedback gains was derived from the covariance dynamics, extending the results of \cite{Moussa2025b} to a more general uncertainty framework. Owing to its reduced dimension, the proposed condition is computationally less demanding than the conventional necessary and sufficient condition, although it may introduce some conservatism. The numerical results validate the exact covariance characterization and compare the two synthesis conditions in terms of computational cost and conservatism.

The proposed deterministic recursive representation of the covariance dynamics may be useful for the development of invariance-based methods, tube-based stochastic MPC, covariance steering, and filtering approaches. Existing necessary and sufficient conditions can be used to design pre-stabilizing gains when computational resources permit. For high-dimensional systems, the proposed sufficient condition provides an alternative that can reduce the computational burden.

Future work will consider time-correlated uncertainties in order to represent a broader class of practical systems. Further directions include the formulation of covariance-steering problems and the use of the proposed characterization for constraint tightening in stochastic tube-based MPC, as well as for invariance- and reachability-based methods.

\section*{Numerical setup}
All computations were performed on a MacBook Pro equipped with an Apple M4 Max chip (16-core CPU, 40-core GPU) and 64 GB of unified memory. The codes were implemented in MATLAB and the LMI conditions were solved using YALMIP \cite{Lofberg2004} and the SDPT3 solver \cite{SDPT32}.

\begin{ack}                              
The work of K.~Moussa was supported by the Clinical project, funded by the ANR under grant ANR-24-CE45-4255.
\end{ack}

\bibliographystyle{unsrt}   
\bibliography{Biblio}   
\end{document}